\documentclass[conference]{IEEEtran}
\ifCLASSINFOpdf
  \usepackage[pdftex]{graphicx}
\else
\fi
\usepackage{amsmath}
\usepackage{url}
\usepackage{tikz}
\usepackage{subfigure}

\usepackage{verbatim}% for multiline comments
\usepackage{booktabs}% http://ctan.org/pkg/booktabs

\begin{document}
%
% paper title
% Titles are generally capitalized except for words such as a, an, and, as,
% at, but, by, for, in, nor, of, on, or, the, to and up, which are usually
% not capitalized unless they are the first or last word of the title.
% Linebreaks \\ can be used within to get better formatting as desired.
% Do not put math or special symbols in the title.
\title{Impact of SDN Controllers Deployment on Network Availability}

\author{\IEEEauthorblockN{Gianfranco Nencioni\IEEEauthorrefmark{1},
Bjarne E. Helvik\IEEEauthorrefmark{1},
Andres J. Gonzalez\IEEEauthorrefmark{2},
Poul E. Heegaard\IEEEauthorrefmark{1} and
Andrzej Kamisi\'{n}ski\IEEEauthorrefmark{3}}
\IEEEauthorblockA{\IEEEauthorrefmark{1}Department of Telematics, Norwegian University of Science and Technology, Trondheim, Norway\\
\{gianfranco.nencioni, bjarne.helvik, poul.heegaard\}@item.ntnu.no}
\IEEEauthorblockA{\IEEEauthorrefmark{2}Telenor Research, Telenor ASA, Trondheim, Norway\\
andres.gonzalez@telenor.com}
\IEEEauthorblockA{\IEEEauthorrefmark{3}Department of Telecommunications, AGH University of Science and Technology, Krak\'{o}w, Poland\\
kamisinski@kt.agh.edu.pl}}

% use for special paper notices
%\IEEEspecialpapernotice{(Invited Paper)}
\IEEEspecialpapernotice{(Technical Report)}

% make the title area
\maketitle

% As a general rule, do not put math, special symbols or citations
% in the abstract
\begin{abstract}
Software-defined networking (SDN) promises to improve the programmability and flexibility of networks, but it may bring also new challenges that need to be explored. The purpose of this technical report is to assess how the deployment of the SDN controllers affects the overall availability of SDN. 
For this, we have varied the number, homing and location of SDN controllers. 
A two-level modelling approach that is used to evaluate the availability of the studied scenarios.
Our results show how network operators can use the approach to find the optimal cost implied by the connectivity of the SDN control platform by keeping high levels of availability.
\end{abstract}

% no keywords

% For peer review papers, you can put extra information on the cover
% page as needed:
% \ifCLASSOPTIONpeerreview
% \begin{center} \bfseries EDICS Category: 3-BBND \end{center}
% \fi
%
% For peerreview papers, this IEEEtran command inserts a page break and
% creates the second title. It will be ignored for other modes.
\IEEEpeerreviewmaketitle

\section{Introduction}\label{sec:Intro}

During the recent years, the SDN has emerged as a new network paradigm, which mainly consists of a programmable network approach where the forwarding plane is decoupled from the control plane \cite{Haleplidis:Software-Defined-Networking-SDN:2015, Kreutz:SDN-survey:2015}.
Despite programmable networks having been studied for decades, 
SDN is experiencing a growing success because it is expected that the ease of changing protocols 
and provide support for adding new services and applications
will foster future network innovation, which is limited and expensive in todays legacy systems.

\begin{figure}[htb]%
\centering 
   \includegraphics[width=\columnwidth]{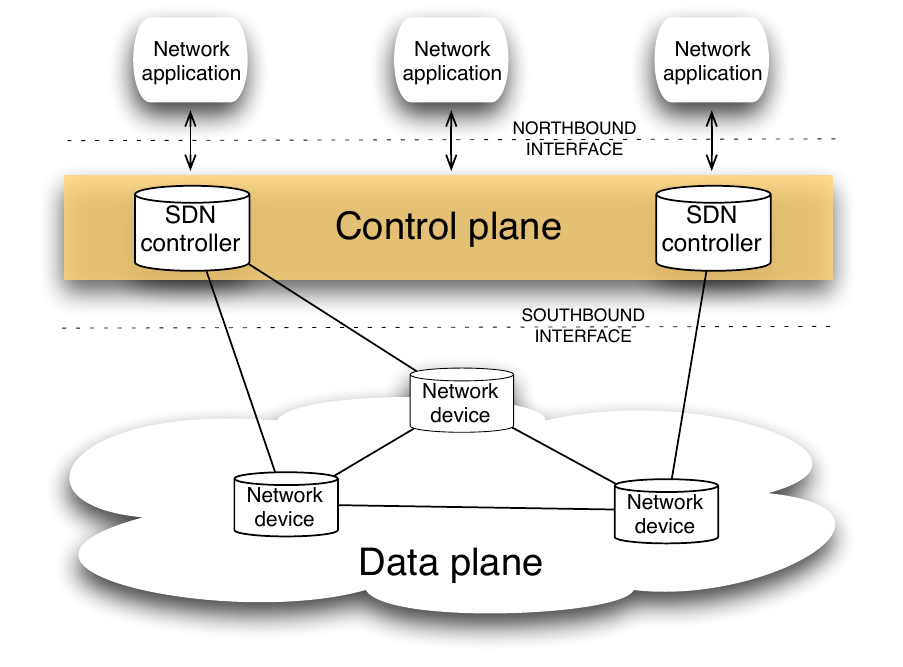} % requires the graphicx package
\caption{SDN architecture (exclusive the management plane)}
\label{fig:SDN-arch}
\end{figure}

A simplified sketch of the SDN architecture from IRFT RFC 7426~\cite{Haleplidis:Software-Defined-Networking-SDN:2015} 
without the management plane is depicted in Figure~\ref{fig:SDN-arch}.
The control plane and data plane are separated.
Here the control plane is logically centralised in a software-based controller (``network brain"), while the data plane is composed of the network devices (``network arms") that conduct the packet forwarding.

The control plane has a northbound and a southbound interface. 
The northbound interface provides an network abstraction to the network applications 
(e.g. routing protocol, firewall, load balancer, anomaly detection, etc...), 
while the southbound interface (e.g. OpenFlow) standardises the information exchange between control and data planes.

In~\cite{Nunes:A-Survey-of-Software-Defined-Networking::2014}, the following set of potential advantages of SDN were pointed out:
\begin{itemize}
 \item centralised control;
 \item simplified algorithms;
 \item commoditising network hardware;
 \item eliminating middle-boxes;
 \item enabling the design and deployment of third-party applications.
\end{itemize}

However, from a dependability perspective, the SDN poses a set of new vulnerabilities and challenges compared with traditional networking, as discussed in~\cite{Heegaard:RNDM:2015}:
\begin{itemize}
 \item consistency of network information (user plane state information) and controller decisions;
 \item consistency between the distributed SDN controllers in the control plane;
 \item increased failure intensities of (commodity) network elements;
 \item compatibility and interoperability between general purpose, non-standard network elements
 \item interdependency between path setup in network elements and monitoring of the data plane in the control plane;
 \item load sharing (to avoid performance bottleneck) and fault tolerance in the control plane have conflicting requirements;
\end{itemize}

The objective of this technical report are to compare the overall availability of SDN when the number, homing and location of SDN controllers are varied.
Note that this work needs to be intent as a preliminary study of \cite{s2DISN}.

that may be achieved with SDN to that of a traditional IP routed network and to investigate under which parametric condition one is better than the other.  In order to do this, we introduce a two level modelling approach, where the top-level catches the structural properties of the networks, and the lower layer the dependability characteristics of the different network elements/subsystems  under hardware, software and operational model.  The models of network elements/subsystems in the two kinds of networks are developed in order to maintain similarities and establish a parametric relation.

In Section~\ref{sec:Model}, we introduce the two-level hierarchical model that has been used in this study.
The evaluation of the deployment of the SDN controllers has reported in Section~\ref{sec:NumEval}.
Finally, the conclusions are summarized in Section~\ref{sec:Concl}.

%%%%%%%%%%%%%%%%%%%%%%%%%%%%%%%%%%%%%%%%%%%%%%%%%%%%%%%%%%%%%%%%%%%%%

\section{Modelling} \label{sec:Model}

A two-levels model (initially introduced in \cite{Heegaard:Trivedi:2016} and then extended \cite{s2DISN}) is used to evaluate the dependability of SDN in a global network.
In particular, the dependability is measured in terms of steady state availability, in the following referred to as availability.
The two-level hierarchical availability modelling approach consists of:
\begin{itemize}
\item {\em Structural} model of the network topology;
\item {\em Dynamic} model of network elements.
\end{itemize}
The approach seeks to avoid the potential uncontrolled growth in model size by compromising the need for modelling details and at the same time modelling a (very) large scale network.
The detailed modelling is necessary to capture the dependencies that exists between network elements, and to described multiple failure modes that might be found in some of the network elements and in the controllers.
The structural model disregards this and assumes independence between the components considered, where a component can be either a single network element with one failure mode, or a set of elements that are interdependent and/or experience several failure modes with an advanced recovery strategy.
For the dynamic models we can use a Markov model or Stochastic Petrinet (e.g., Stochastic Reward Network~\cite{Ciardo:Trivedi:93}). For the structural model we can use reliability block diagram, fault trees, or structure functions based on minimal cut or path sets.

The objective of the modelling approach is to compare the availability of SDN with a traditional IP network 
with the same topology of network elements (SDN forwarding switched and IP routers).

%%%%%%%%%%%%%%%%%%%%%%%%%%%%%%%%%%%%%%%%%%%%%%%%%%%%%%%%%%%%%%%%%%%%%

\section{Numerical evaluation} \label{sec:NumEval}

\begin{figure}[htbp]
\begin{center}
\includegraphics[width=\columnwidth]{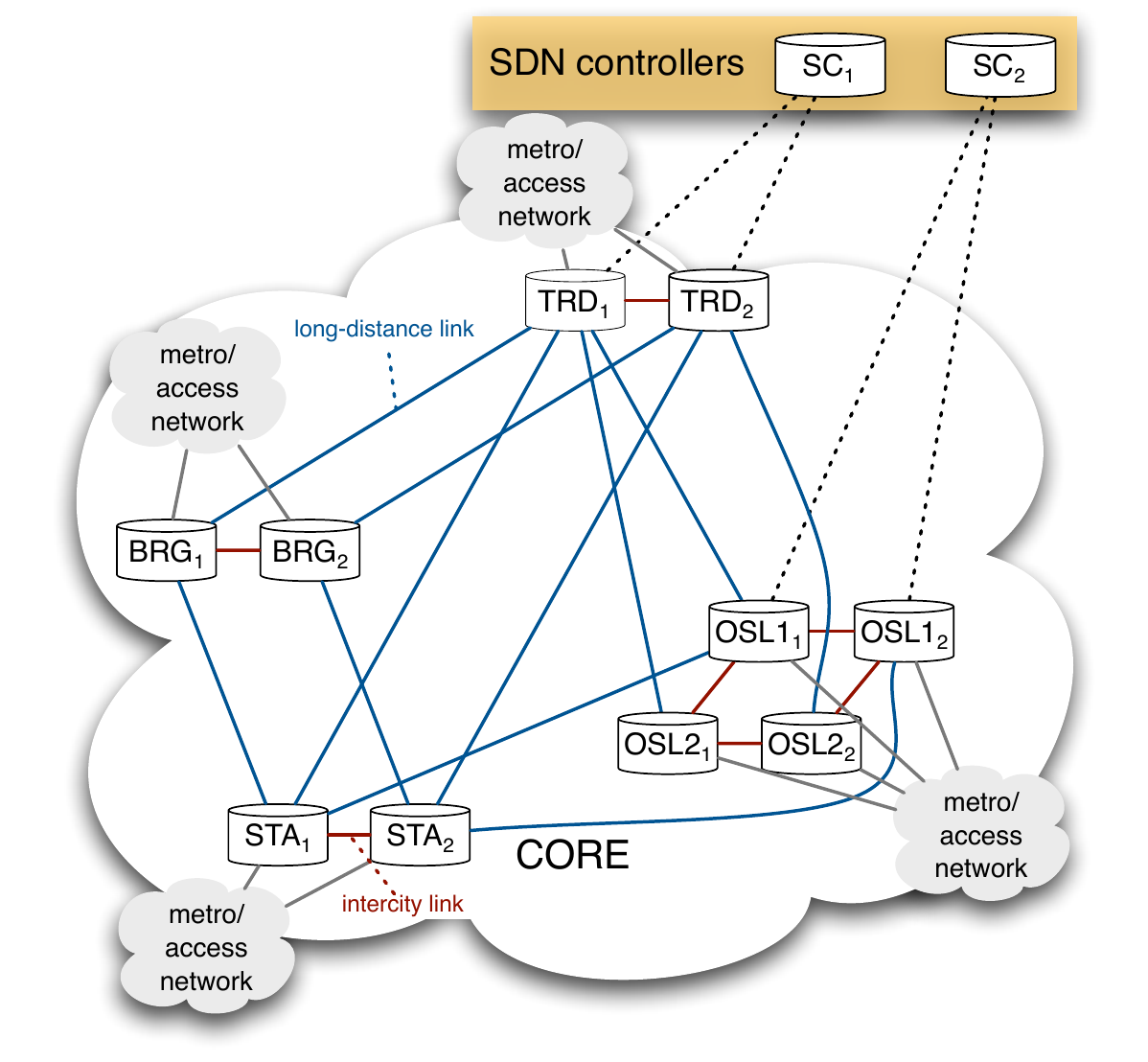}
\caption{Topology of the nation-wide backbone network}
\label{fig:SDN-nation-backbone}
\end{center}
\end{figure}

In this evaluation we consider the national backbone network depicted in Figure~\ref{fig:SDN-nation-backbone} and consists of 10 nodes across 4 cities, 
and two dual-homed SDN controllers.
The nodes are located in the four major cities in Norway,  Bergen (BRG), Trondheim (TRD), Stavanger (STV), and Oslo (OSL). 
Each town has duplicated nodes, except Oslo which has four nodes (OSL1 and OSL2).
The duplicated nodes are labelled, $X_1$ and $X_2$, where $X$=OSL1, OSL2, BRG, STV, and TRD. 
In addition to the forwarding nodes, there are two dual-homed SDN controllers (SC$_1$ and SC$_2$), which are connected to TRD and OSL1. 

We assume that nodes, links, and controllers in the system may fail. 
The peering traffic in a city is routed through an access and metro network with a connection to both (all four) nodes in the city.
The system is working (up), when all the access and metro networks are connected. 
Note that for SDN, at least one controller must be reachable from all nodes along a working path.

To evaluate the availability of traditional networks and SDN, we consider the same typical parameters used in \cite{s2DISN}, 
which are inspired by and taken from several studies~\cite{Gonzalez:Characterization-of-Router-and-Link:2012, Kuusela:On/off-process-modeling:2010, Verbrugge:General-availability-model:2005}.

\begin{figure*}
 \centering
 \subfigure[Case 1]
  {\includegraphics[width=.8\columnwidth]{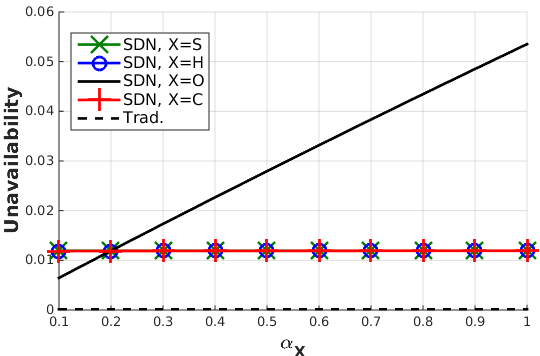}\label{fig:UAconn_1}} 
 \subfigure[Case 2]
  {
  \includegraphics[width=.8\columnwidth]{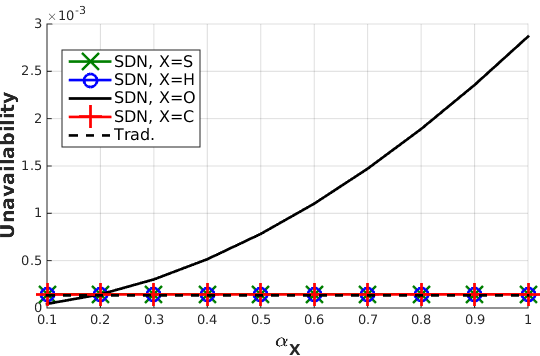}\label{fig:UAconn_2}} 
 \subfigure[Case 3]
  {   \includegraphics[width=.8\columnwidth]{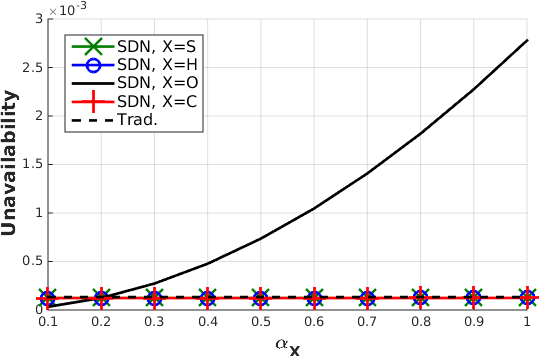}\label{fig:UAconn_3}} 
 \subfigure[Case 4]
  {\includegraphics[width=.8\columnwidth]{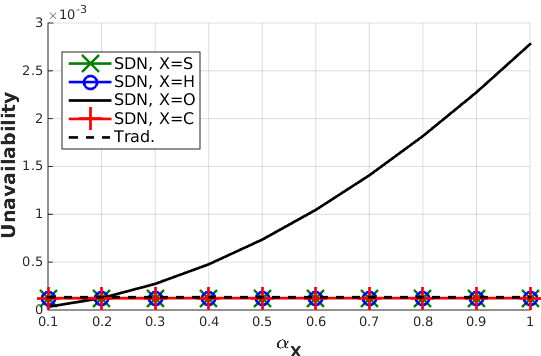}\label{fig:UAconn_4}} 
 \subfigure[Case 5]
  {\includegraphics[width=.8\columnwidth]{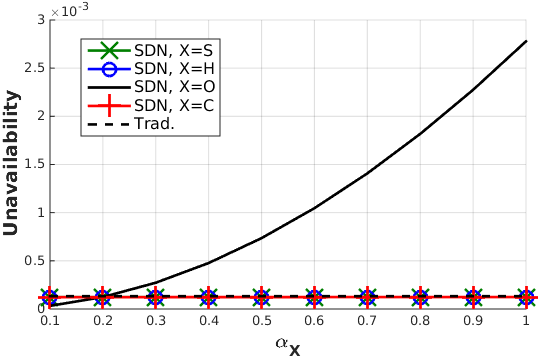}\label{fig:UAconn_5}} 
 \subfigure[Case 6]
  {\includegraphics[width=.8\columnwidth]{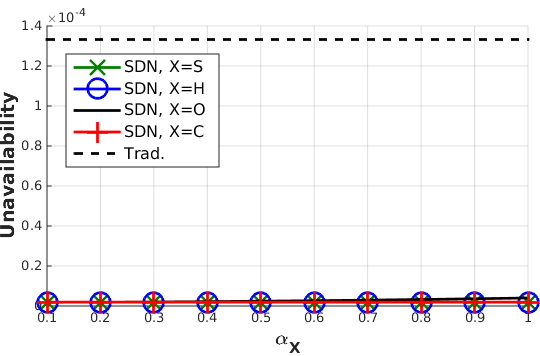}\label{fig:UAconn_6}} 
 \subfigure[Case 7]
  {\includegraphics[width=.8\columnwidth]{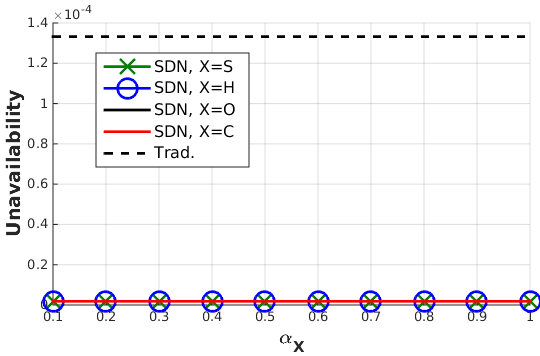}\label{fig:UAconn_7}} 
 \subfigure[Case 8]
  {\includegraphics[width=.8\columnwidth]{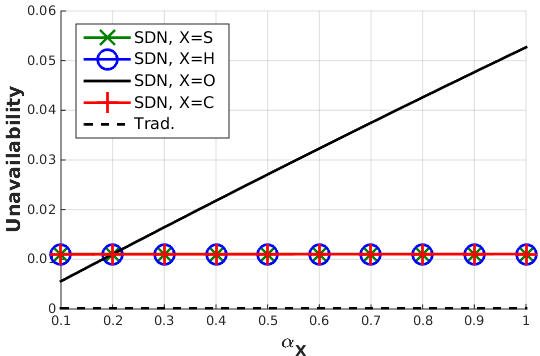}\label{fig:UAconn_8}} 
 \caption{Unavailability by varying $\alpha$s in the different connectivity and redundancy configurations}
 \label{fig:UAconn_tot}
\end{figure*}

To evaluate the impact of the connectivity and the redundancy of the SDN controllers in the national backbone network we consider the following case studies:
\begin{enumerate}
 \item There is only one controller and it is connected to $OSL1_2$;
 \item There are two single-homed controllers ($SC1$ connected to $TRD_2$ and $SC2$ connected to $OSL1_2$);
 \item Reference scenario depicted in Figure~\ref{fig:SDN-nation-backbone};
 \item The controllers are triple-homed (added connections from $SC1$ to $BRG_1$ and from $SC2$ to $BRG_2$ to the reference scenario);
 \item The controllers are quadruple-homed (added connections from $SC1$ to $STV_1$ and from $SC2$ to $STV_2$ to the previous scenario);
 \item There are three controllers (added controller connected to $BRG_1$ and to $BRG_2$ to the reference scenario);
 \item There are four controllers (added controller connected to $STV_1$ and to $STV_2$ to the previous scenario); 
 \item There is one dual-homed controller (deleted $SC2$ from the reference scenario). 
\end{enumerate}

Figure~\ref{fig:UAconn_tot} shows the unavailability of SDN in the case studies.
In the figure the $\alpha_X$ factors where $X=S,H,O,C$, which affect the intensity of the related failure sources (software, $\alpha_S$, hardware, $\alpha_H$, O\&M, $\alpha_O$, and coverage, $\alpha_C$) and are defined as follows:
\begin{itemize}
 \item $\alpha_H=\dfrac{\lambda_H}{N/K~\lambda_{dC}}$;
 \item $\alpha_S=\dfrac{\lambda_S}{N~\lambda_{dS}}$;
 \item $\alpha_O=\dfrac{\lambda_O}{N~\lambda_{dO}}$.
\end{itemize}

\subsection{Evaluating SDN controller connectivity}

To evaluate the impact of the SDN controllers connectivity in the national backbone network we consider the case studies 2, 3, 4, and 5.

\begin{figure}
 \centering
 \subfigure[Case 2]
  {\includegraphics[width=.8\columnwidth]{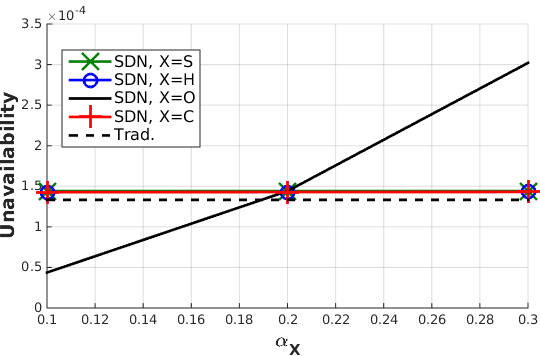}} 
 \subfigure[Case 3]
  {\includegraphics[width=.8\columnwidth]{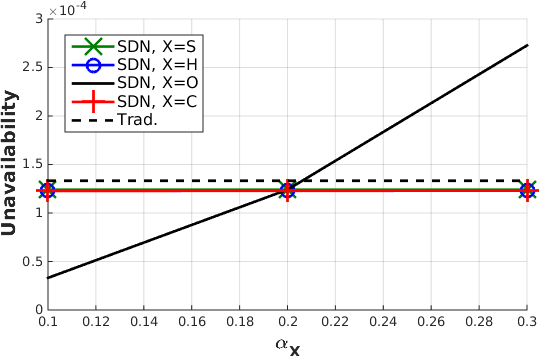}} 
 \caption{Unavailability by varying $\alpha$s in the different connectivity configurations (zoomed version)}
 \label{fig:UAconn_zoom} 
\end{figure}

Figures \ref{fig:UAconn_3}, \ref{fig:UAconn_4}, and \ref{fig:UAconn_5} highlight that the unavailability in the cases 3, 4, and 5 is almost the same, so having a triple- or quadruple-homed controller would not enhance the availability performance but would increase the deployment cost especially if inter-city connections are needed. 
The most critical case is of course the case 1 when there is just one single-homed controller, in this case the unavailability in increased by one order of magnitude.
The difference between cases 2 and 3 (better depicted in Figure~\ref{fig:UAconn_zoom}) is really small, this is likely due to the high availability of the links.

\subsection{Evaluating SDN controller redundancy}

To evaluate the impact of the SDN controllers redundancy in the national backbone network we consider the case studies 3, 6, 7, and 8.

Figures \ref{fig:UAconn_3} and \ref{fig:UAconn_8} show the network availability in the case with two dual-homed controllers and the case with one dual-homed controller. the comparison highlights that the unavailability with one controller is from one to two orders of magnitude higher than the unavailability with two controller and two orders magnitude higher that the unavailability of the traditional network.

\begin{figure}
 \centering
 \subfigure[Case 6]
  {\includegraphics[width=.8\columnwidth]{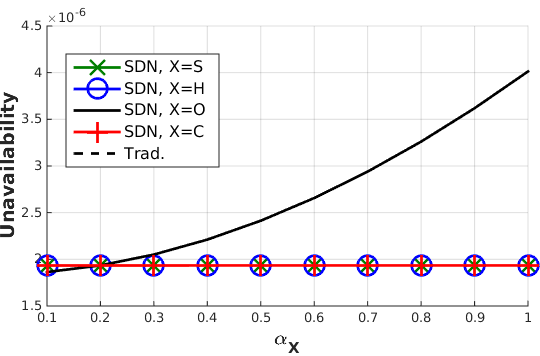}} 
 \subfigure[Case 7]
  {\includegraphics[width=.8\columnwidth]{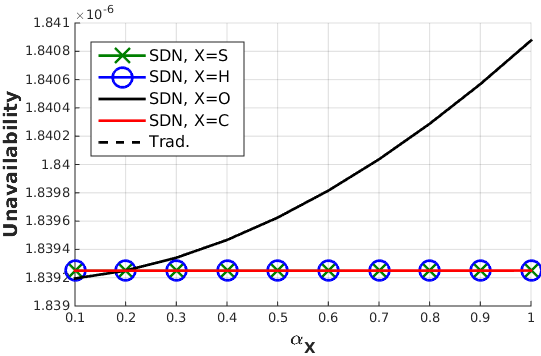}} 
 \caption{Unavailability by varying $\alpha$s in the different redundancy configurations (zoomed version)}
 \label{fig:UAredun_zoom} 
\end{figure}

Figures \ref{fig:UAconn_6} and \ref{fig:UAconn_7} show that the unavailability in the cases of three and four dual-homed controllers is lower than the unavailability of the traditional network. Figure \ref{fig:UAredun_zoom} highlights that the unavailability with three and four controllers is on the same order of magnitude and two orders of magnitude lower than both traditional network and the SDN with two controllers.

\subsection{Evaluating SDN controller location}

\begin{figure}[htbp]
\begin{center}
\includegraphics[width=.95\columnwidth]{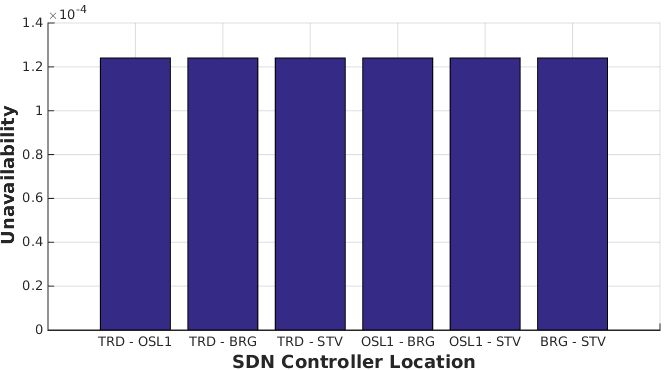}
\caption{Unavailability by varying the location of the SDN controllers ($\alpha_S=1$, $\alpha_H=1$,  $\alpha_O=0.2$, and $\alpha_C=1$)}
\label{fig:SDNlocation}
\end{center}
\end{figure}

Figure~\ref{fig:SDNlocation} shows that the unavailability of SDN does not change by varying the location of the SDN controller. We note only a minor increase when one of the controller is connected to OSL1, we suppose that is because OSL1 belongs to Oslo that is the only city with 4 nodes.
Anyway, the problem of SDN controller placement should be more deeply analysed in a larger and more complex network.

%%%%%%%%%%%%%%%%%%%%%%%%%%%%%%%%%%%%%%%%%%%%%%%%%%%%%%%%%%%%%%%%%%%%%

\section{Conclusions}  \label{sec:Concl}
In this technical report, we have evaluated how the overall availability of SDN is influenced by the number, homing and location of SDN controllers.
The results show that having a triple- or quadruple-homed controller would not enhance the availability performance, but would unnecessary increase the deployment cost to be made by operators.
Single-homing has reduced the availability as wall but the impact is limited.
Comparing the number of SDN controllers, one SDN controller is not enough to guarantee an acceptable availability. Two controllers allows the SDN to have an availability comparable to the traditional network. Instead, three and four controller permit to obtain a really low unavailability but the deployment cost may be excessive to a network operator, which target is to provide an availability similar to the current traditional network.
In the addressed study, the location of the SDN controllers has not influence on the overall availability.

\newpage

% trigger a \newpage just before the given reference
% number - used to balance the columns on the last page
% adjust value as needed - may need to be readjusted if
% the document is modified later
%\IEEEtriggeratref{2}
% The "triggered" command can be changed if desired:
%\IEEEtriggercmd{\enlargethispage{-5in}}

% references section

% can use a bibliography generated by BibTeX as a .bbl file
% BibTeX documentation can be easily obtained at:
% http://mirror.ctan.org/biblio/bibtex/contrib/doc/
% The IEEEtran BibTeX style support page is at:
% http://www.michaelshell.org/tex/ieeetran/bibtex/
\bibliographystyle{IEEEtran}
% argument is your BibTeX string definitions and bibliography database(s)
\bibliography{reference}

% Generated by IEEEtran.bst, version: 1.13 (2008/09/30)
\begin{thebibliography}{10}
\providecommand{\url}[1]{#1}
\csname url@samestyle\endcsname
\providecommand{\newblock}{\relax}
\providecommand{\bibinfo}[2]{#2}
\providecommand{\BIBentrySTDinterwordspacing}{\spaceskip=0pt\relax}
\providecommand{\BIBentryALTinterwordstretchfactor}{4}
\providecommand{\BIBentryALTinterwordspacing}{\spaceskip=\fontdimen2\font plus
\BIBentryALTinterwordstretchfactor\fontdimen3\font minus
  \fontdimen4\font\relax}
\providecommand{\BIBforeignlanguage}[2]{{%
\expandafter\ifx\csname l@#1\endcsname\relax
\typeout{** WARNING: IEEEtran.bst: No hyphenation pattern has been}%
\typeout{** loaded for the language `#1'. Using the pattern for}%
\typeout{** the default language instead.}%
\else
\language=\csname l@#1\endcsname
\fi
#2}}
\providecommand{\BIBdecl}{\relax}
\BIBdecl

\bibitem{Haleplidis:Software-Defined-Networking-SDN:2015}
E.~Haleplidis, K.~Pentikousis, S.~Denazis, J.~H. Salim, D.~Meyer, and
  O.~Koufopavlou, ``Software-defined networking ({SDN}): Layers and
  architecture terminology,'' Internet Research Task Force (IRTF), Request for
  Comments RFC 7426, January 2015.

\bibitem{Kreutz:SDN-survey:2015}
D.~Kreutz, F.~M.~V. Ramos, P.~J.~E. Ver{\'{\i}}ssimo, C.~E. Rothenberg,
  S.~Azodolmolky, and S.~Uhlig, ``Software-defined networking: {A}
  comprehensive survey,'' \emph{Proceedings of the {IEEE}}, vol. 103, no.~1,
  pp. 14--76, 2015.

\bibitem{Nunes:A-Survey-of-Software-Defined-Networking::2014}
B.~Nunes, M.~Mendonca, X.-N. Nguyen, K.~Obraczka, and T.~Turletti, ``A survey
  of software-defined networking: Past, present, and future of programmable
  networks,'' \emph{Communications Surveys Tutorials, IEEE}, vol.~16, no.~3,
  pp. 1617--1634, Third 2014.

\bibitem{Heegaard:RNDM:2015}
P.~E. Heegaard, B.~E. Helvik, and V.~B. Mendiratta, ``Achieving dependability
  in software-defined networking - a perspective,'' in \emph{7th International
  Workshop on Reliable Networks Design and Modeling (RNDM'15)}, M{\"u}nich,
  Germany, October 5-7 2015, p.~7.

\bibitem{s2DISN}
G.~Nencioni, B.~E. Helvik, A.~J. Gonzalez, P.~E. Heegaard, and A.~Kami\'{n}ski,
  ``{Availability Modelling of Software-Defined Backbone Networks},''
  \emph{submitted to the 2nd Workshop on Dependability Issues on SDN and NFV
  (DISN 2016)}.

\bibitem{Heegaard:Trivedi:2016}
P.~E. Heegaard, B.~E. Helvik, G.~Nencioni, and J.~W{\"a}fler, ``Managed
  dependability in interacting systems,'' in \emph{Principles of Performance
  and Reliability Modeling and Evaluation}, L.~Fiondella and A.~Puliafito,
  Eds.\hskip 1em plus 0.5em minus 0.4em\relax Springer, 2016.

\bibitem{Ciardo:Trivedi:93}
G.~Ciardo and K.~S. Trivedi, ``A decomposition approach for stochastic reward
  net models,'' \emph{Perf. Eval}, vol.~18, pp. 37--59, 1993.

\bibitem{Gonzalez:Characterization-of-Router-and-Link:2012}
A.~J. Gonzalez and B.~E. Helvik, ``Characterization of router and link failure
  processes in {UNINETT}'s {IP} backbone network,'' \emph{International Journal
  of Space-Based and Situated Computing}, 2012.

\bibitem{Kuusela:On/off-process-modeling:2010}
P.~Kuusela and I.~Norros, ``On/off process modeling of ip network failures,''
  in \emph{Dependable Systems and Networks (DSN), 2010 IEEE/IFIP International
  Conference on}, June 2010, pp. 585--594.

\bibitem{Verbrugge:General-availability-model:2005}
S.~Verbrugge, D.~Colle, P.~Demeester, R.~Huelsermann, and M.~Jaeger, ``General
  availability model for multilayer transport networks,'' in
  \emph{Proceedings.5th International Workshop on Design of Reliable
  Communication Networks, 2005. (DRCN 2005)}.\hskip 1em plus 0.5em minus
  0.4em\relax IEEE, October 16-19 2005, pp. 85 -- 92.

\end{thebibliography}
%
% <OR> manually copy in the resultant .bbl file
% set second argument of \begin to the number of references
% (used to reserve space for the reference number labels box)
% \begin{thebibliography}{1}
% 
% \bibitem{IEEEhowto:kopka}
% H.~Kopka and P.~W. Daly, \emph{A Guide to \LaTeX}, 3rd~ed.\hskip 1em plus
%   0.5em minus 0.4em\relax Harlow, England: Addison-Wesley, 1999.
% 
% \end{thebibliography}

% that's all folks
\end{document}